\documentclass[prb,twocolumn,showpacs,amsmath,amssymb]{revtex4-1}

\usepackage[dvipdfmx]{graphicx}
\usepackage{bm}
    \usepackage{color}
\usepackage{braket}

\begin{document}

\title{Generalized Bloch band theory for non-Hermitian bulk-boundary correspondence}



\author{Ken-Ichiro Imura}
\author{Yositake Takane}

\affiliation{Graduate School of Advanced Science and Engineering, Hiroshima University, 739-8530, Japan}

\date{\today}

\begin{abstract}
Bulk-boundary correspondence is the cornerstone of topological physics.
In some non-Hermitian topological systems
this fundamental relation is broken
in the sense that 
the topological number calculated for the Bloch energy band
under the periodic boundary condition
fails to reproduce the boundary properties under the open boundary.
To restore the bulk-boundary correspondence in such non-Hermitian systems
a framework beyond the Bloch band theory is needed.
We develop a non-Hermitian Bloch band theory based 
on a modified periodic boundary condition
that allows a proper description of the bulk 
of a non-Hermitian topological insulator
in a manner
consistent with its boundary properties.
Taking a non-Hermitian version of the 
Su-Schrieffer-Heeger 
model as an example,
we demonstrate our scenario,
in which
the concept of bulk-boundary correspondence is naturally generalized
to non-Hermitian topological systems.
\end{abstract}

\maketitle

\section{Introduction}


In topological insulators
\cite{KM1,KM2,SCZ1,SCZ2,HasanKane,HG,ryu}
non-trivial topology of the bulk wave function
is unambiguously related to the appearance of mid-gap boundary states.
This relation often referred to as bulk-boundary correspondence\cite{HG,ryu}
highlights the physics of topological insulator.
However, it has been pointed out that
this concept cannot be trivially applied to
some non-Hermitian topological insulators.\cite{YaoWang}
In these systems
Bloch band theory 
fails to encode nontrivial boundary 
properties
in the bulk energy band,\cite{YM}
thus breaking superficially the bulk-boundary correspondence.
To restore the bulk-boundary correspondence 
a framework beyond the standard Bloch band theory is needed.

Making quantum mechanics non-Hermitian 
might have been regarded as an act of heresy. 
\cite{NH1,NH2}
The parity-time symmetric non-Hermitian systems with real eigenvalues 
\cite{Bender}
might have been less guilty; 
indeed such systems have been much studied.
\cite{PT_0,PT_1,PT_2,PT_3,PT_4,PT_5,hide_1,topo_4}
But after a decade or so,
non-Hermitian quantum mechanics is now regarded as real.
\cite{topo_2,silicon,nphys_QW,photo,nature,science}
The non-Hermitian generalization of a topological insulator
has also started to be considered in this context
\cite{Msato,hughes}.
Since then, an increasing number of papers have appeared
in this field,
\cite{tony,YaoWang,YM,Longhi,Slager,anom_loc,biortho,why,svd,transfer_M,gil,thomale,Foa_Torres,Chen_Fang,Okuma_PRL,Lee1,Lee2}
including a remarkable proposal of
a practical application, the topological insulator laser.
\cite{SSH_GL_laser,topo_3,topo_1,TI_laser2}

A nontrivial feature of a topological insulator is in
its energy spectrum under an open boundary
condition (obc).
Let us consider, for simplicity,
a one-dimensional model of topological insulator of length $L$.
If $L$ is sufficiently large,
the energy eigenvalues arrange themselves into
quasi continuous energy bands,
which we will simply call {\it bulk energy bands}.\cite{YM}
Such bulk energy bands also allows us to introduce an energy gap and 
the concept of an insulator.
For the system to be topologically non-trivial, 
there must appear boundary states in the enegy gap,
and 
such boundary states must be protected by a bulk topological number.
In an infinite Hermitian crystal
such a topological number is expressed as an integral 
with respect to $k$ over the entire Brillouin zone ($k\in [0,2\pi]$),\cite{kohmoto1, kohmoto2}
where
$k$ specifies a Bloch function:
\begin{equation}
\psi (x) = e^{ikx} u(x).
\label{bloch1}
\end{equation}
$u(x)$ is its periodic part satisfying $u(x+1)=u(x)$,
and the lattice constant $a$ is chosen to be unity ($a=1$).

Figure 1(a) [top annex panel] shows typical energy spectra of a non-Hermitian topological system [Su-Schrieffer-Heeger (SSH) model with non-reciprocal hopping].
The blue spectrum represents the one
in the boundary geometry; i.e., under the obc,
where
contributions from the bulk and from the boundary
combine to give the full spectrum.
Between the two gap closing points $P_1$ and $P_2$ there appears a pair of mid-gap boundary states.
The gray spectrum represents the one
in the bulk geometry
under the periodic boundary condition (pbc).
Note that
in the Hermitian limit
the bulk geometry is equivalent to the Bloch band theory
under the pbc.
Here, in a non-Hermitian example, the number and positions of the gap closing points $Q_{1-4}$
in the bulk geometry (gray/pbc spectrum) are different
from the ones in the boundary geometry (blue/obc spectrum),
clearly indicating that the bulk energy band in the Bloch band theory
based on the pbc
fails to achieve the bulk-boundary correspondence.

\begin{figure*}
(a)
\includegraphics[width=75mm, bb=0 0 354 482]{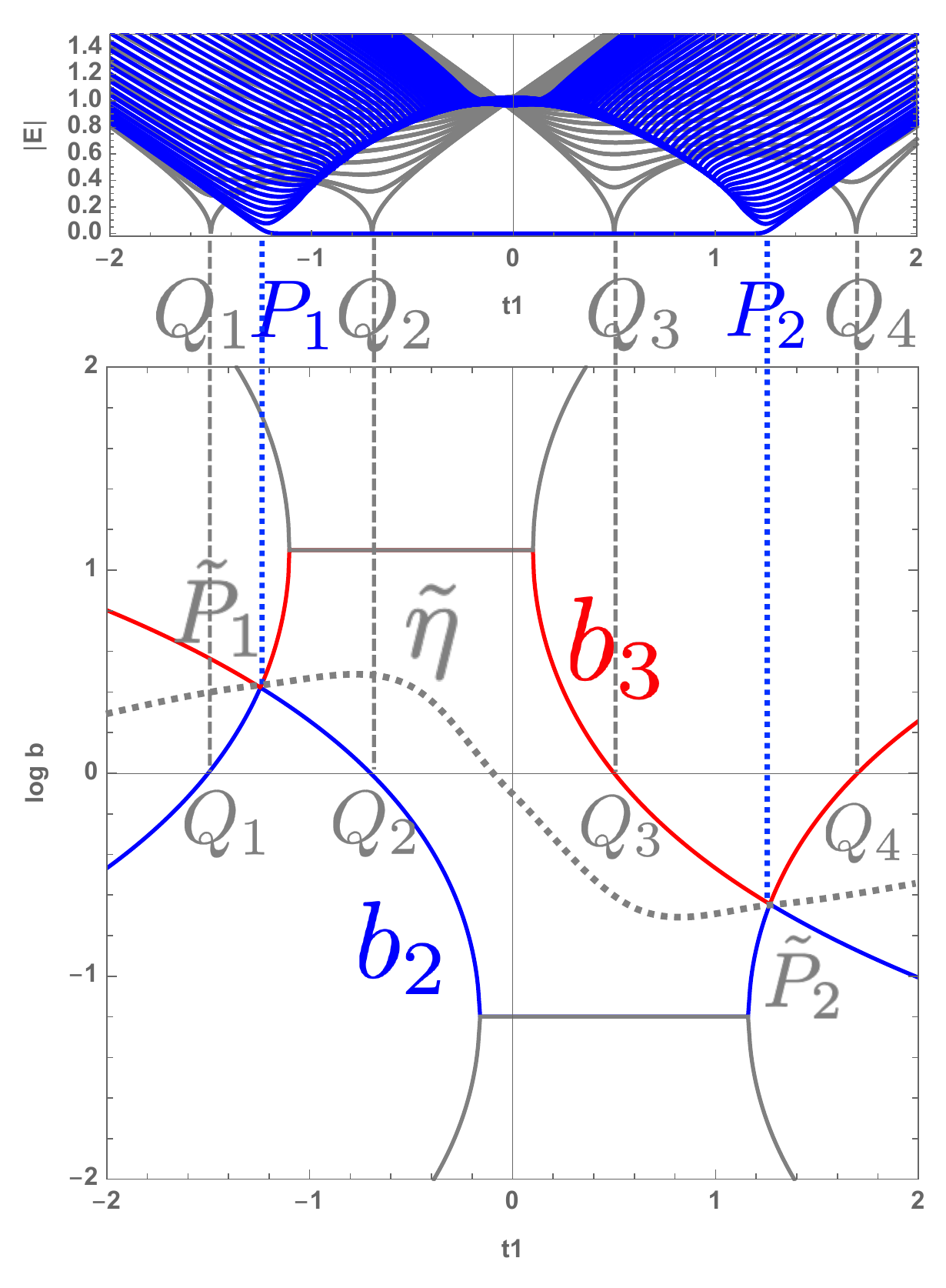}
(b)
\includegraphics[width=75mm, bb=0 0 284 284]{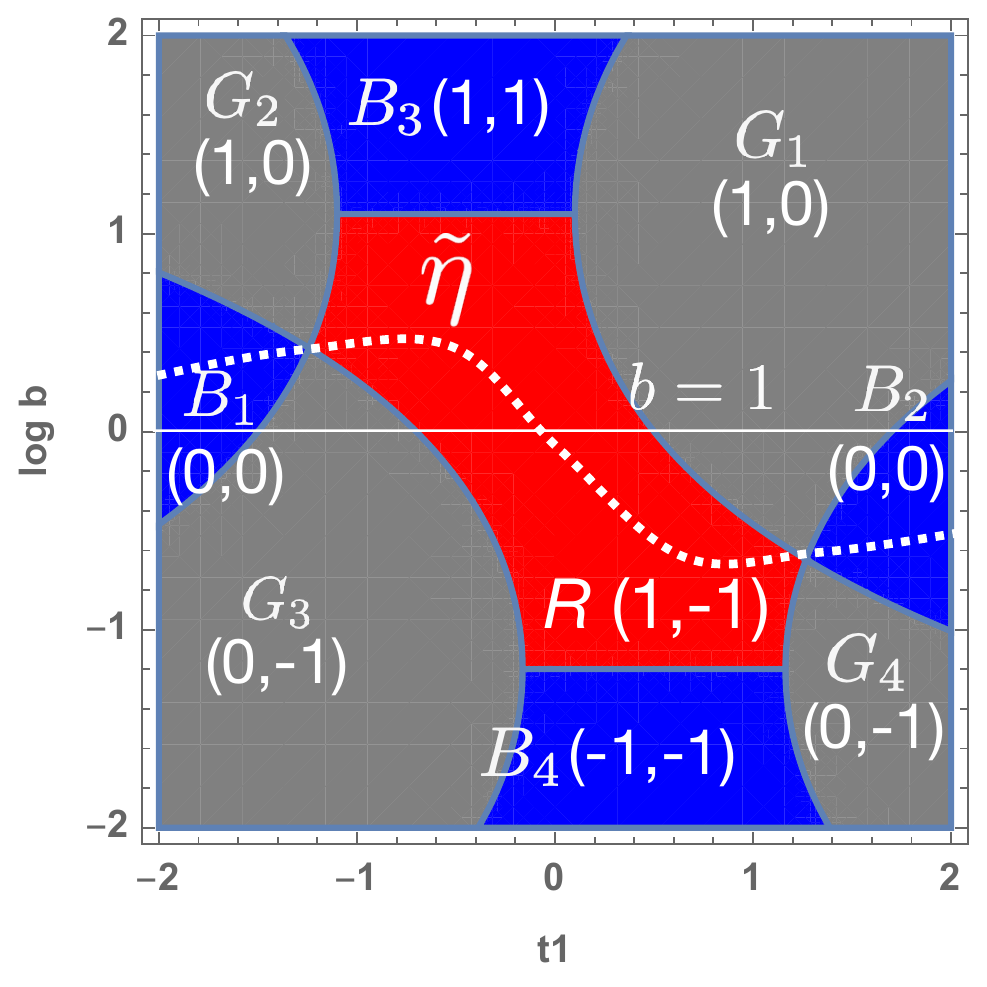}

\vspace{-2mm}
\caption{Schematic illustration of the non-Hermitian bulk-boundary correspondence.
The two main panels show how to decode
the topological information encoded in the spectra of a non-Hermitian SSH model
[extra panel on top of panel (a),
the blue (gray) curves represent the spectrum
under the open (periodic) boundary condition].
The model is explicitly given in Eqs. (\ref{H_NN}), (\ref{t12}) and (\ref{H_3NN}).
Parameters: $t_2=1$, $t_3=0.1$, $\gamma_1=0.5$, $\gamma_2=0.1$.
}
\label{schema}
\end{figure*}



In the above example of a non-Hermitian topological insulator,\cite{YaoWang,YM}
the bulk energy bands under the obc are very different from the ones
in the Bloch band theory under the pbc.
Under the obc
the corresponding eigen wave function is not a superposition of
the plane-wave Bloch form (\ref{bloch1}),
but rather of the following generalized Bloch form:
\begin{equation}
\psi (x) = \beta^x u(x).
\label{bloch2}
\end{equation}
Unless $|\beta| = 1$, 
such a wave function damps (or amplifies) exponentially, 
and tends to be localized in the vicinity of an end of the system.
This phenomenon is often referred to as non-Hermitian skin effect,
and typical to systems with non-reciprocal hopping.
\cite{tony,YaoWang,YM,Longhi,anom_loc,biortho,why,svd,transfer_M,gil,thomale}
Note that Eq. (\ref{bloch2}) is clearly incompatible with  
the pbc: $\psi (x+L)=\psi(x)$ unless $|\beta|=1$.

We have previously established
\cite{IT19}
that by considering a modified periodic boundary condition (mpbc):
$\psi (x+L)=b^L \psi (x)$
with $b$ being a real positive constant, less restrictive than the standard 
pbc,
one can restore the desired correspondence between
the boundary and the bulk properties.
In this paper, we perform a detailed description of the
bulk of this system in this scenario,
which was lacking in Ref. \onlinecite{IT19},
leading to 
a non-Hermitian generalization of the Bloch band theory
compatible with the bulk-boundary correspondence scenario.

The argument of Ref. \onlinecite{IT19} on the bulk-boundary correspondence
was based on a smooth deformation to the Hermitian limit.
Here,
we give a more general argument 
supporting that
the bulk-boundary correspondence proposed in Ref. \onlinecite{IT19}
is valid also in the non-perturbative non-Hermitian regime.
To demonstrate our scenario
we consider a one-dimensional ($d=1$) model;
e.g., Hatano-Nelson model and a non-Hermitian generalization of the SSH model.
Yet, the underlying idea can be equally applied to systems with higher spatial dimensions
($d=2,3$).
Our approach can be also applied to other class of models,
e.g., for a  system with symplectic symmetry. \cite{KK_PRB2020,Okuma_PRL}
The authors of Refs. \onlinecite{YaoWang,YM} 
proposed an alternative scenario that extracts the information on the boundary properties from the bulk energy band under the obc (i.e., the continuum part of the obc spectrum).
This scenario takes only the boundary geometry into consideration.
We, on the contrary, employ the genuine bulk geometry under the mpbc to extract 
the information on the boundary properties. 
Unlike the non-Bloch approach of Refs. \onlinecite{YaoWang,YM},
our scenario may be regarded as a natural generalization of the 
Bloch band theory to non-Hermitian system.
Note that our generalized Bloch band theory 
includes the conventional Hermitian Bloch band theory
as a special case of $b=1$.

\section{Hermitian and Non-Hermitian Bloch band theories}

\subsection{Generalized Bloch function under the mpbc}

Let us consider the eigenstates of an electron in a crystal.
Let the Hamiltonian $H$ possess a lattice translational symmetry;
$H$ commutes with
the lattice translation operator $\Xi$ such that
$[\Xi, H]=0$,
where
$\Xi$ acts on the eigenstate (eigen bra) $\langle x|$ 
of the coordinate $x$ such that
$\langle x|\Xi= \langle x+1|$.
Here, we have chosen the lattice constant $a$ to be unity ($a=1$).
Let $|\beta\rangle$ be a simultaneous eigenstate of
$H$ and $\Xi$,
\begin{eqnarray}
H |\beta\rangle &=& E(\beta) |\beta\rangle,
\nonumber \\
\Xi |\beta\rangle &=& \beta |\beta\rangle.
\label{beta_eigen}
\end{eqnarray}
The wave function defined by
$\psi (x) = \langle x |\beta\rangle$
satisfies the relation:
\begin{equation}
\psi (x+1) = \langle x |\Xi |\beta\rangle = \beta \langle x |\beta\rangle =\beta\psi (x),
\label{x+1}
\end{equation}
implying that
$\psi (x)$ takes the generalized Bloch form (\ref{bloch2}).
In an infinite crystal the wave function $\psi (x)$ must be bounded wherever in the system 
governed by Hermitian quantum mechanics.
This imposes $|\beta|=1$, or $\beta=e^{ik}$ with $k \in [0,2\pi]$, and 
hence 
$\psi (x)$
reduces to the standard Bloch form (\ref{bloch1}).
When $\beta$ takes continuous values on a unit circle in the complex 
$\beta$-pane,
the trajectory of $E(\beta)$ defines a Bloch energy band.
Thus, the range of $k \in [0,2\pi]$, or equivalently, the unit circle in 
the $\beta$-plane defines the Brillouin zone (BZ) for a Hermitian crystal.

Applying a pbc is standard in the conventional Bloch band theory
\footnote{J. M. Ziman, {\it Principles of the Theory of Solids}, Cambridge University Press, 1972 (Second edition).}
and allows one to reconstitute the above continuous energy band,
starting with a system of finite length $L$
and taking the limit $L\rightarrow\infty$ at the end of the calculation.
The pbc: $\psi (x+L)=\psi (x)$
imposes $\beta^L=1$,
i.e.,
$\beta=e^{ik}$ with $k=2\pi n/L$, $n=0,1,2,\cdots,L-1$.
Namely,
the pbc, 
without making reference to the boundedness of the wave function,
automatically selects from the generalized Bloch form (\ref{bloch2}),
those wave functions that have the standard plane-wave like form (\ref{bloch1})
with $|\beta|=1$.

The pbc is technically important 
for an unambiguous formulation of the Bloch band theory,
avoiding difficulties arising from the continuous values of $k$ in an infinite system.
In Hermitian topological band insulators
the pbc plays the role of bulk geometry,
in which purely bulk states appear in the spectrum.
The existence and identification of
such a geometry is indispensable for achieving the bulk-boundary correspondence.
However, in non-Hermitian systems
the pbc fails to play a proper role of such a bulk geometry.

\begin{figure}
\centering\includegraphics[width=80mm, bb=0 0 425 142]{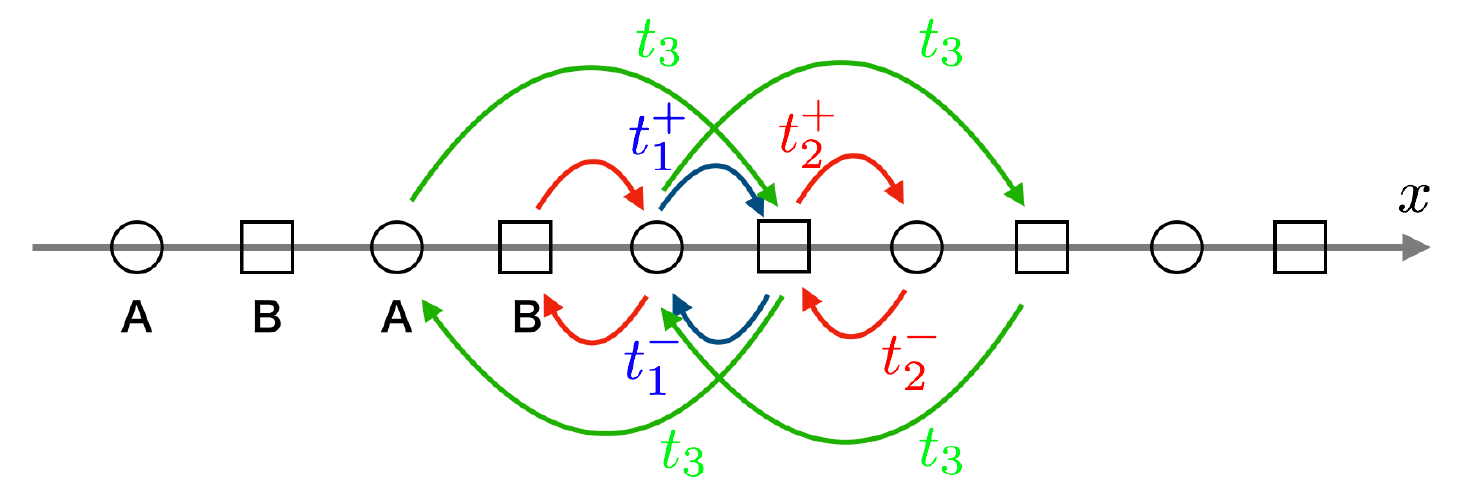}
\vspace{-2mm}
\caption{A non-Hermitian extension
of the SSH (Su-Schriffer-Heeger) model employed in Refs. \onlinecite{YaoWang,YM}. 
Here, we have updated the convention of $t_1^\pm$ and $t_2^\pm$.
}
\label{model}
\end{figure}

\subsection{Implementation of the mpbc to a tight-binding model}

Let us consider the Hatano-Nelson model \cite{NH1,NH2} in the clean limit:
\begin{equation}
H_{HN} =\sum_{x} (t_R |x+1\rangle\langle x| + t_L |x\rangle\langle x+1|).
\label{HN_infini}
\end{equation}
The hopping amplitudes $t_R$ and $t_L$ are chosen to be non-reciprocal: $t_R\neq t_L$.
The recipe for imposing the standard pbc: $\psi (x+L)=\psi (x)$
is well known;
one truncates the summation over $x$ in
Eq. (\ref{HN_infini}) as 
\begin{equation}
H_{HN}^{(L)} =\sum_{x=1}^{L-1} (t_R |x+1\rangle\langle x| + t_L |x\rangle\langle x+1|).
\label{HN_L}
\end{equation}
and
adds the boundary term:
\begin{equation}
H_{bd}(1) = t_R |1\rangle\langle L| + t_L |L\rangle\langle 1|,
\label{H_bd1}
\end{equation}
which couples the last site ($x=L$)
back to the first one ($x=1$)
via non-reciprocal hopping terms.
The total Hamiltonian
$H_{pbc}^{HN}=H_{HN}^{(L)}+H_{bd}(1)$
represents the system under the standard pbc.
The eigenstates of $H_{pbc}^{HN}$ take 
the following standard Bloch form:
\begin{equation}
|\beta\rangle=\sum_{x=1}^L \beta^x |x\rangle,
\label{beta_L}
\end{equation}
with $\beta^L=1$, i.e., $|\beta|=1$.

However, Eq. (\ref{H_bd1}) is not a unique choice to make the system periodic by 
closing it via a boundary hopping term.
Instead of Eq. (\ref{H_bd1}), one can equally consider 
\begin{equation}
H_{bd}(b) = b^{-L} t_R |1\rangle\langle L| + b^L t_L |L\rangle\langle 1|,
\label{H_bd2}
\end{equation}
where $b$ is a real positive constant.
\cite{IT19}
The form of the boundary term (\ref{H_bd2}) implies 
that the eigenstates of 
$H_{mpbc}^{HN}(b)=H_{HN}^{(L)}+H_{bd}(b)$
are in the form of Eq. (\ref{beta_L}) with $\beta$ satisfying $\beta^L=b^L$,
i.e., $|\beta|=b$ or $\beta=be^{ik}$.
The eigen wave function $\psi(x)$ satisfies the mpbc:\cite{IT19}
\begin{equation}
\psi (x+L)=b^L \psi (x)
\label{mpbc_wf}
\end{equation}
At this stage we do not attempt to determine 
the parameter $b$, but keep it as a free parameter in the theory. 
For a given $b$, our mpbc specifies that
the allowed values of $\beta$
are on a circle of radius $b$ in the complex $\beta$-plane,
which represents the generalized Brillouin zone
of the system.
$b$ can be chosen, for example, as $b=\sqrt{t_R/t_L}$ 
representing the degree of skin effect under the obc (see below),
but if one's purpose is simply to restore the bulk-boundary correspondence,
it has been shown that
such a fine tuning of $b$ is unnecessary except at the bulk gap closing.\cite{IT19}

In an infinite system specified by Eq. (\ref{HN_infini}),
\begin{equation}
|\beta\rangle=\sum_{x} \beta^x |x\rangle
\label{beta_infini}
\end{equation}
with an arbitrary $\beta$ 
becomes formally an eigenstate of the system,
although
such a wave function is generally
unbounded and not compatible with quantum mechanics.
The corresponding eigenenergy $E(\beta)$ is determined as
\begin{equation}
t_R\beta^{-1}+t_L\beta=E=E(\beta).
\label{sec_HN}
\end{equation}
From these infinite number of eigenstates,
the boundary Hamiltonian (\ref{H_bd2})
selects those solutions satisfying $\beta^L=b^L$.
As for the standard pbc, it selects only those solutions with 
$\beta^L=1$.
This turns out to be too restrictive and inadequate
for a description of the bulk
in the bulk-boundary correspondence.

Let us comment on the obc case, in which the Schr\"odinger equation is
$H_{HN}^{(L)}|\psi_{obc}\rangle=E|\psi_{obc}\rangle$ with
$|\psi_{obc}\rangle =\sum_x\psi_{obc}(x) |x\rangle$.
The wave function $\psi_{obc}(x) $ subjected to the boundary condition of $\psi_{obc}(0) = 0$ is expressed as a superposition of the generalized Bloch form (\ref{beta_L}):
\begin{equation}
\psi_{obc}(x)\propto \beta_1^x-\beta_2^x,
\label{wf_obc}
\end{equation}
where $\beta_1$ and $\beta_2$ are the two solutions of Eq. (\ref{sec_HN}) at an equal energy $E$,
which implies $\beta_1\beta_2=t_R/t_L\neq 1$.
We also need to impose the boundary condition: $\psi_{obc}(L+1)=0$.
This enforces 
\begin{equation}
(\beta_1/\beta_2)^{L+1}=1,
\label{obc_L}
\end{equation}
implying together with the previous condition,
$|\beta_1|=|\beta_2|=\sqrt{t_R/t_L} \neq 1$.
As a result 
$\psi_{obc}(x)$
tends to damp or amplify exponentially towards the end of the system
(non-Hermitian skin effect).\cite{YaoWang}

\section{A non-Hermitian SSH model}

We present a non-Hermitian extension of the Su-Schriffer-Heeger (SSH) model \cite{ssh}
(see Fig.~\ref{model}), which is used to illustrate our scenario in 
subsequent sections.
Among different extensions
\cite{simon}
we consider here a variation with non-reciprocal hopping,
since
this type of non-Hermiticity is more problematic for the bulk-boundary correspondence
than the complex potentials (gain/loss type non-Hermiticity).
\cite{tony,bipolar,gil,thomale,YaoWang,YM}

\subsection{The model}

A non-Hermitian version of the SSH model:
\begin{eqnarray}
H_{\rm NN}^{(L)}
&=& \sum_{x=1}^L
\left[t_1^+ |x,B\rangle\langle x,A| + t_1^- |x,A\rangle\langle x,B|
\right]
\nonumber \\
&+&
\sum_{x=1}^{L-1}
\left[
t_2^+
|x+1,A\rangle\langle x,B| + 
t_2^-
|x,B\rangle\langle x+1,A|\right],
\label{H_NN}
\end{eqnarray}
gives a prototypical example,
in which one encounters the difficulty of applying the standard pbc
to a non-Hermitian system.
\cite{tony}
The nearest-neighbor (NN) hopping amplitudes:
\begin{equation}
t_1^\pm=t_1\pm\gamma_1,\ \
t_2^\pm=t_2\pm\gamma_2,
\label{t12}
\end{equation}
are chosen to be non-reciprocal.
$t_{1}^\pm$
represents an intra-cell hopping amplitudes in the $\pm x$ direction.
$t_{2}^\pm$
represents an inter-cell hopping amplitudes in the $\pm x$ direction.
Here, 
following Refs. \onlinecite{YaoWang,YM}
we also consider third-nearest-neighbor (3NN) 
hopping  terms $t_3$ such that
\begin{equation}
H_{\rm 3NN}^{(L)}
= \sum_{x=1}^{L-1}
\left[t_3 |x+1,B\rangle\langle x,A| + t_3 |x,A\rangle \langle x+1,B|
\right].
\label{H_3NN}
\end{equation}
Our total Hamiltonian reads
\begin{eqnarray}
H_{\rm SSH}^{(L)}&=&H_{\rm NN}^{(L)}+H_{\rm 3NN}^{(L)}.
\label{SSH_obc}
\end{eqnarray}
We assume that the model parameters
$t_\mu (\mu=1,2,3)$,
$\gamma_\mu (\mu=1,2)$
are all real constants.
The Hamiltonian (\ref{SSH_obc}) with Eqs. (\ref{H_NN}) and (\ref{H_3NN})
describes our system
under the obc.


\subsection{The mpbc: the boundary Hamiltonian and the generalized Bloch Hamiltonian}

To discuss bulk-boundary correspondence in the non-Hermitian SSH model (\ref{SSH_obc})
we need to consider it in a bulk geometry.
This is realized by introducing boundary hopping terms that close the system.
We introduce boundary terms that concretize the 
mpbc,
which have NN and 3NN parts:
\begin{equation}
H_{bd}^{\rm SSH}(b)=H_{bd}^{\rm NN}(b)+H_{bd}^{\rm 3NN}(b),
\end{equation}
where
\begin{eqnarray}
H_{bd}^{\rm NN}(b)
&=& 
b^{-L} t_2^+
|1,A\rangle\langle L,B| + 
b^L t_2^-
|L,B\rangle\langle 1,A|,
\nonumber \\
H_{bd}^{\rm 3NN}(b)
&=& 
b^{-L} t_3 |1,B\rangle\langle L,A| + 
b^L t_3 |L,A\rangle \langle 1,B|.
\label{bd_mpbc}
\end{eqnarray}
These hopping matrix elements 
couple the final unit cell $x=L$ 
back to the first one $x=1$ (and vice versa),
with an amplitude amplification ($b^L$) or attenuation ($b^{-L}$) factor
to select eigenstates of an appropriate amplitude $b$ (see Sec. II).
Note also that
the boundary Hamiltonian (\ref{bd_mpbc})
is non-Hermitian unless $b=1$.
The total Hamiltonian under the mpbc reads
\begin{equation}
H_{mpbc}^{\rm SSH}(b)=H_{\rm SSH}^{(L)}+H_{bd}^{\rm SSH}(b).
\label{SSH_mpbc}
\end{equation}

\begin{figure*}
(a)\includegraphics[width=45mm, bb=0 0 284 284]{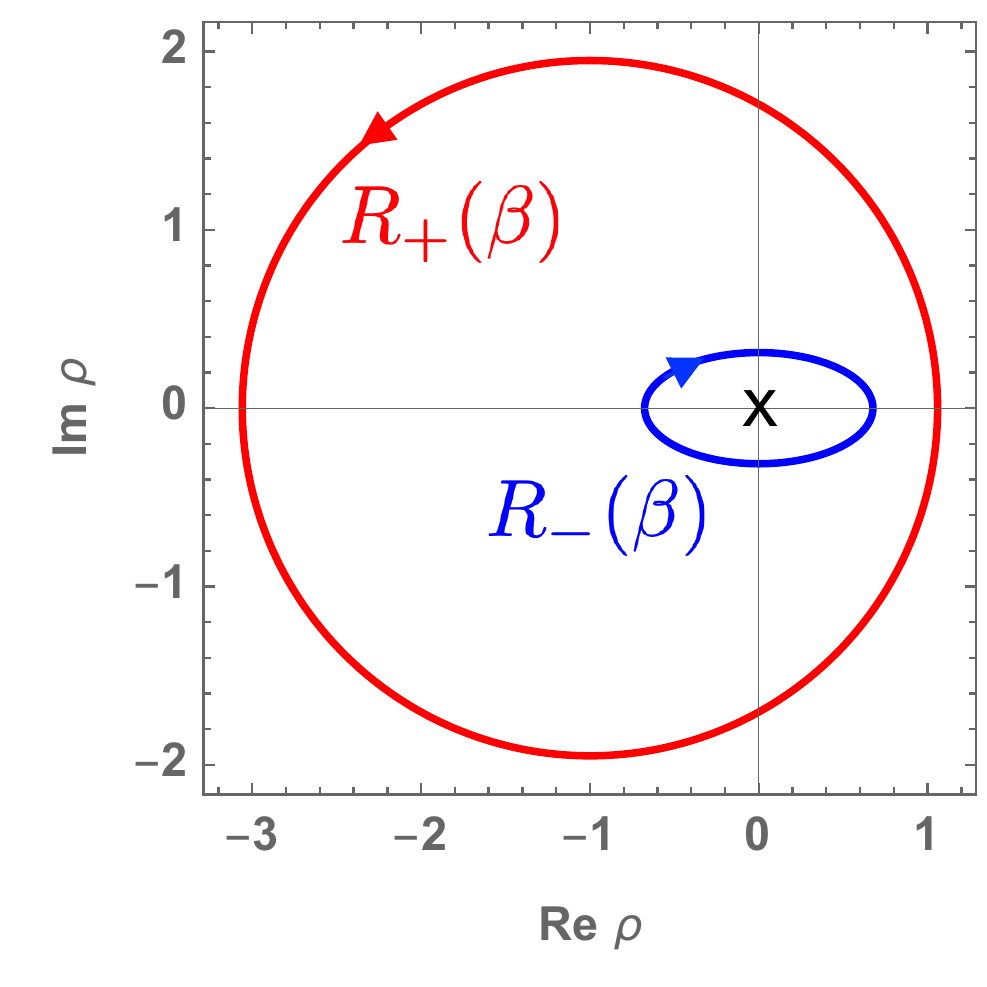}
(b)\includegraphics[width=45mm, bb=0 0 284 284]{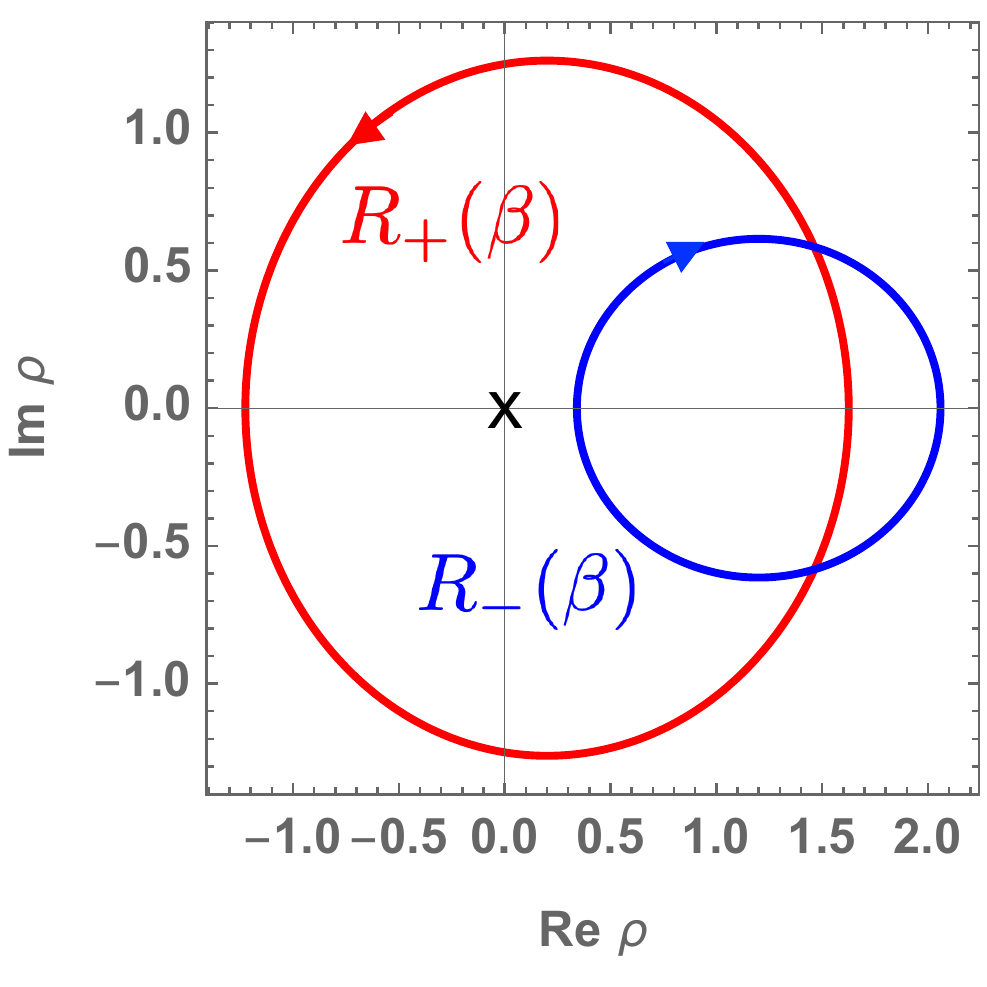}
(c)\includegraphics[width=45mm, bb=0 0 284 284]{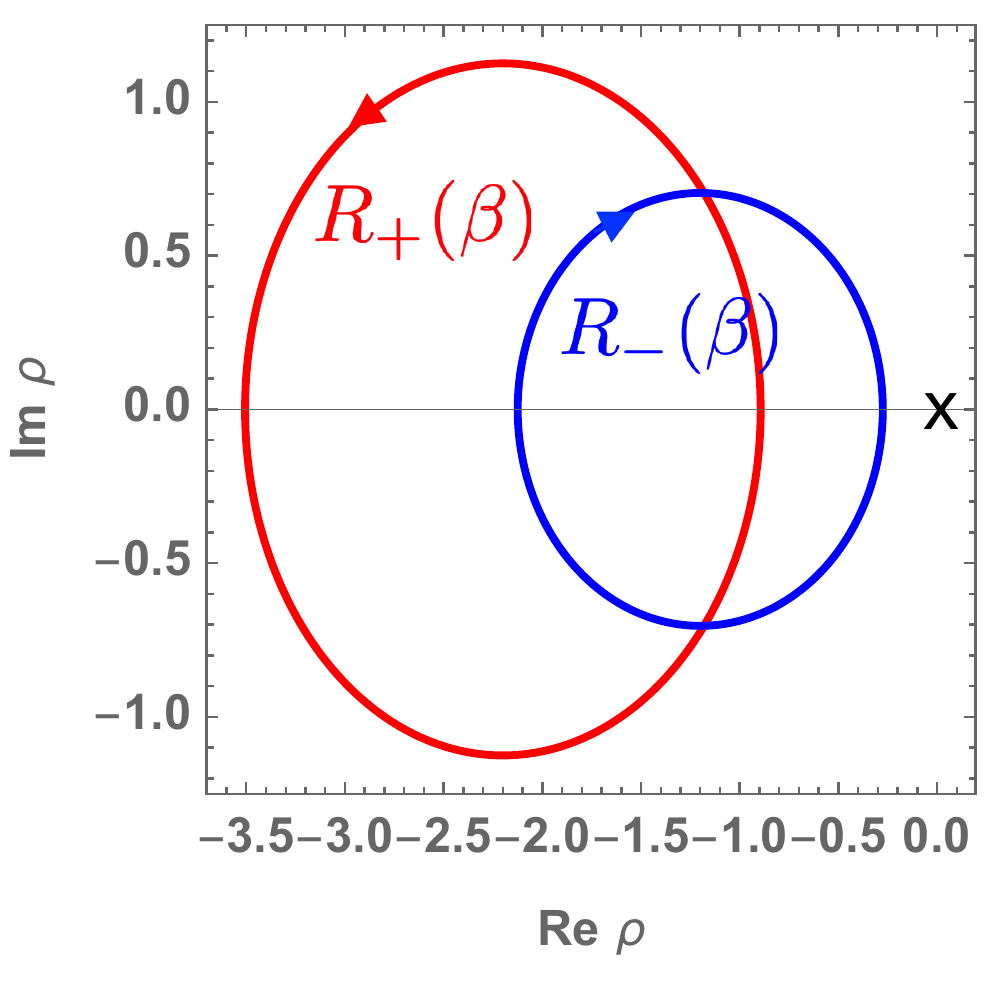}
\vspace{-2mm}
\caption{
The trajectory of $\rho_\pm=R_\pm (\beta)$ [cf. Eq. (\ref{rho_pm})]
at different points in Fig. 1
with $t_2=1$, $t_3=0.1$, $\gamma_1=0.5$, $\gamma_2=0.1$,
where $\times$ indicates the position of origin.
Panel (a): $(t_1,b)=(-0.5, 0.7)$ in the $R$ region (TI phase) with
$(w_+,w_-)=(1,-1)$ and $w=1$.
Panel (b): $(t_1,b)=(0.6, 0.2)$ in the $G_1$ region with
$(w_+,w_-)=(1,0)$ and $w=1/2$.
Panel (c): $(t_1,b)=(-1.7, 0.1)$ in the $B_1$ region (OI phase) with
$(w_+,w_-)=(0,0)$ and $w=0$.
}
\label{trajec_pm}
\end{figure*}

As in the single-band case,
the eigenstate of $H_{mpbc}^{\rm SSH}(b)$
takes a generalized Bloch form:
\begin{equation}
|\beta\rangle = \sum_{x=1}^L \beta^x (u_A |x,A\rangle
+u_B|x,B\rangle),
\label{beta2}
\end{equation}
where
$\beta^L=b^L$,
i.e.,
$\beta=be^{ik}$ with $k=2\pi n/L$, $n=0,1,2,\cdots,L-1$.
Such values of $\beta$ 
are on a circle of radius $b$ in the complex $\beta$-plane,
which represents our generalized Brillouin zone in the limit of $L\rightarrow\infty$.
Equation (\ref{beta2}) is a two-band generalization of Eq. (\ref{beta_L}).
Then, 
the eigenvalue problem:
\begin{equation}
H_{mpbc}^{\rm SSH} |\beta\rangle = E(\beta)|\beta\rangle,
\label{SSH_eigen}
\end{equation}
reduces to the following $2\times 2$ problem
specified by the generalized Bloch Hamiltonian $H_{mpbc}(\beta)$:
\begin{eqnarray}
H_{mpbc}(\beta)
\left[\begin{array}{c}
u_A\\u_B
\end{array}\right]
=E(\beta)
\left[\begin{array}{c}
u_A\\u_B
\end{array}\right],
\end{eqnarray}
where 
\begin{equation}
H_{mpbc}(\beta)=
\left[
 \begin{array}{cc}
0&R_- (\beta)\\
 R_+ (\beta) &0
  \end{array}
\right],
\label{H_bloch}
\end{equation}
and
\begin{eqnarray}
R_+(\beta) &=& t_1^+ + t_2^-\beta + t_3 \beta^{-1},
\nonumber \\
R_-(\beta) &=& t_1^- + t_2^+\beta^{-1} + t_3 \beta.
\label{R_pm}
\end{eqnarray}
Note that
$H_{mpbc}(\beta)$ is chiral (sublattice \cite{KK_PRX}) symmetric
so that our system belongs to class AIII.
\cite{KK_PRX,kitaev,schnyder1,schnyder2}

For a given $\beta=be^{ik}$
the corresponding energy eigenvalue $E(\beta)$ is determined by the
secular equation:
\begin{equation}
R_+(\beta) R_-(\beta)=E^2,
\label{quartic}
\end{equation}
which is a quartic equation for $\beta$ in the present case.
Note that generally $E$ is complex.

\subsection{Case of the obc}

To achieve the bulk-boundary correspondence in the present model
one needs to consider also its boundary geometry.\cite{ryu,IT19}
The boundary geometry is realized by imposing the obc. 
As noted before, this is equivalent to consider $H_{\rm SSH}^{(L)}$ as the total Hamiltonian of the system.
Thus by diagonalizing 
$H_{\rm SSH}^{(L)}$ one finds the obc spectrum.
When the model parameters are those corresponding to the topologically nontrivial phase
[topological insulator (TI) phase],
one finds a pair of boundary states at $E=0$ in the spectrum. 
The rest of the spectrum forms asymptotically 
a bulk energy band
in the limit of $L\rightarrow\infty$.
If $E$ falls on 
the bulk energy band,
the corresponding eigenstate $|E_{obc}\rangle$
under the obc 
must have the following characteristics.\cite{YM,YaoWang}

In the present model
the secular equation (\ref{quartic})
is quartic so that
for a given $E$ there are four solutions $\beta_j$s ($j=1,2,3,4$) for $\beta$.
Unlike under the mpbc,
the amplitudes $|\beta|$ are not {\it a priori} specified quantities.
We label $\beta_j$s in the increasing order of their magnitude such that
\begin{equation}
|\beta_1|\le|\beta_2|\le|\beta_3|\le|\beta_4|.
\label{order}
\end{equation}
The eigenstate $|E_{obc}\rangle$ under the obc 
can be expressed as a superposition of four fundamental solutions 
such that
\begin{equation}
|E_{obc}\rangle
=c_1|\beta_1\rangle
+c_2|\beta_2\rangle
+c_3|\beta_3\rangle
+c_4|\beta_4\rangle,
\label{psi4}
\end{equation}
where $|\beta_j\rangle$s are given as
\begin{equation}
|\beta_j\rangle = \sum_{x} \beta_j^x (u_A |x,A\rangle
+u_B|x,B\rangle)
\label{beta2_infini}
\end{equation}
for $j=1,2,3,4$.
Each $|\beta_j\rangle$ is a fundamental solution of the infinite system
and not an eigenstate compatible with the obc.
The two central components of Eq. (\ref{psi4}) are 
analogues of the plane-wave components $e^{ikx}$ and $e^{-ikx}$ in the Hermitian case,
while the other components associated with
$\beta_1$ and $\beta_4$ are sorts of evanescent modes 
to make the function to be compatible with the obc.
For the wave function (\ref{psi4}) to be an eigenstate in the bulk
energy band under the obc, $|\beta_2\rangle$ and $|\beta_{3}\rangle$ must satisfy\cite{YaoWang,YM}
\begin{equation}
|\beta_2| = |\beta_3|
\label{YM_cond}
\end{equation}
in the limit of $L\rightarrow\infty$.
This condition, in turn, determines the allowed values of $E$ forming 
the bulk energy band.
The trajectory of $\beta_2$ or $\beta_{3}$
satisfying Eq. (\ref{YM_cond})
defines the generalized Brillouin zone
of this obc approach.\cite{YaoWang,YM}
Note that under the mpbc we employ a different generalized Brillouin zone.

\section{Non-Hermitian winding numbers}

As noted earlier our one-dimensional model is chiral symmetric.
In the Hermitian limit, 
it is well established 
\footnote{J. K. Asboth, L. Oroszlany, and A. Palyi, A Short Course on Topological Insulators: Band Structure and Edge States in One and Two Dimensions, Lecture Notes in Physics Vol. 919 (Springer, Berlin, 2016).}
that such a system is classified by a $\mathbb Z$-type topological number, or a winding number.
\cite{kitaev,schnyder1,schnyder2}

\subsection{A pair of non-Hermitian chiral winding numbers} 

Extrapolating our knowledge in the Hermitian limit,
let us consider the following pair of chiral winding numbers
defined in the bulk geometry under the mpbc:
\begin{eqnarray}
     w_\pm(b)&=& {1\over 2\pi}[ \arg R_\pm (\beta) ]_{k=0}^{2\pi}
                    \nonumber \\
             &=& {1\over 2\pi}[\phi_\pm (2\pi)-\phi_\pm (0)],
\label{w_pm}
\end{eqnarray}
where $\beta=be^{ik}$, $R_\pm$ are given in Eqs. (\ref{R_pm}), and
\begin{equation}
\phi_\pm (k) = {\rm Im} \log R_\pm (\beta).
\end{equation}
Here, for $k$ to be continuous,
the limit $L\rightarrow\infty$ is implicit.
The winding numbers $w_\pm(b)$ are functions of
model parameters $t_\mu$, $\gamma_\mu$
and also of $b$ specifying the mpbc. 
In this sense $w_\pm$ are defined in the generalized parameter space 
\begin{equation}
\tilde{\tau}=\{b,\tau\},
\end{equation}
where
\begin{equation}
\tau=\{t_\mu, \gamma_\mu\}
\end{equation}
is the original set of parameters.
$w_\pm$ measures how many times
the trajectory,
\begin{equation}
  \rho=R_\pm(\beta) \equiv\rho_\pm
  \label{rho_pm}
\end{equation}
winds around the origin 
in the complex $\rho$-plane
while $\beta=be^{ik}$ goes around the generalized Brillouin zone
once in the anti-clockwise direction
(while $k$ sweeps the standard BZ: $k\in[0,2\pi]$ once from 0 to $2\pi$).
The three panels of Fig. \ref{trajec_pm}
represent the trajectory $\rho_\pm(t_1,b)$ 
at different points in Fig. 1 (the winding number map).

In the Hermitian limit: $\gamma_1=\gamma_2=0$, $b=1$,
the Hermiticity requires $R_- = R_+^*$,
so that $w_-$ is fixed to $- w_+$.
The winding number $w$ is defined as
\begin{equation}
w \equiv {w_+ - w_- \over 2}=w_+=-w_-.
\label{sym}
\end{equation}
It is well established that $w=1$ encodes
the existence of a pair of boundary states at $E=0$ (TI phase),
while $w=0$ encodes the absence of boundary states 
[ordinary insulator (OI) phase],
giving a prototypical 
example of the bulk-boundary correspondence.

\begin{figure*}
(a)\includegraphics[width=45mm, bb=0 0 284 284]{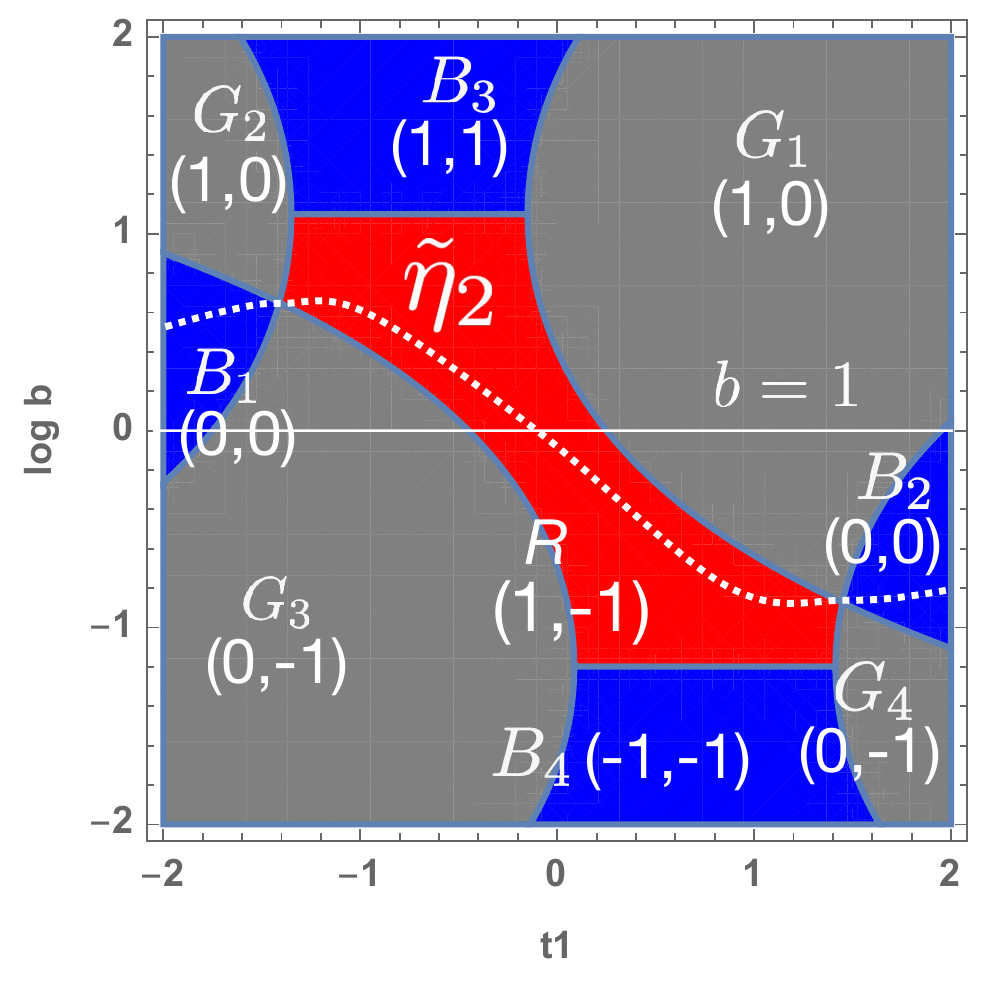}
(b)\includegraphics[width=45mm, bb=0 0 284 284]{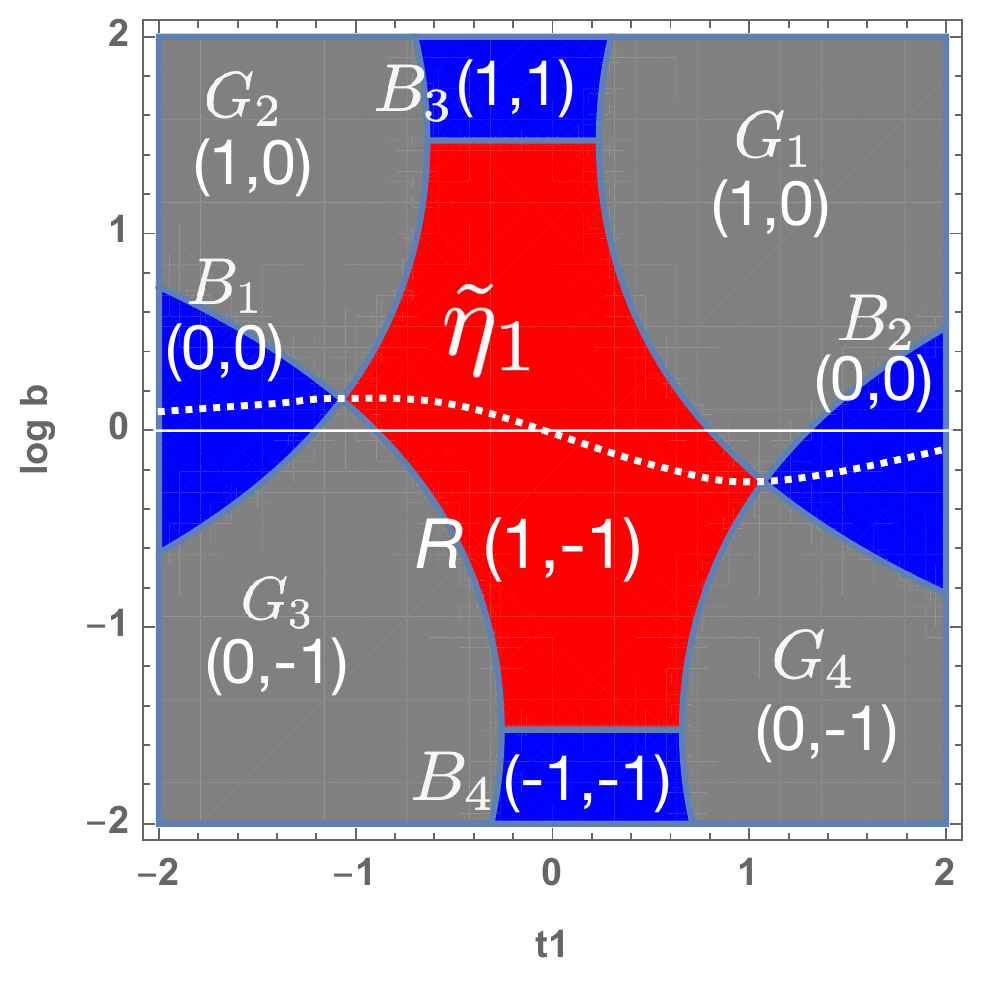}
(c)\includegraphics[width=45mm, bb=0 0 284 284]{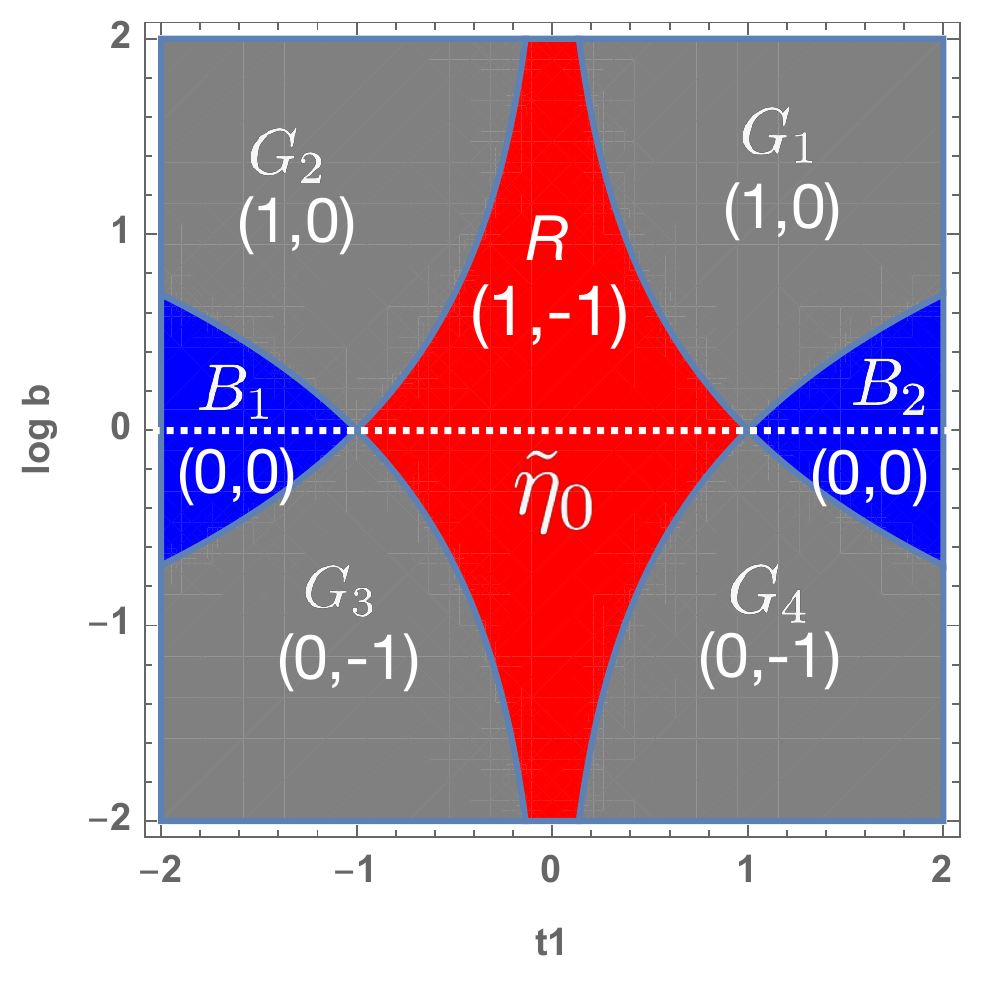}
\vspace{-2mm}
\caption{Evolution of the path $\tilde{\eta}$ in the perturbative non-Hermitian regime. Panel (a) describes a situation analogous to the one in Fig. 1 (non-Hermiticity is slightly strong): $t_2=1$, $t_3=0.05$, $\gamma_1=0.75$, $\gamma_2=0.1$. Panel (c) describes the Hermitian limit: $t_2=1$, $t_3=\gamma_1=\gamma_2=0$. Panel (b) corresponds to an intermediate situation with $t_2=1$, $t_3=0.05$, $\gamma_1=0.2$, $\gamma_2=0.05$. The path $\tilde{\eta}$ evolves as
$\tilde{\eta}_2\rightarrow\tilde{\eta}_1\rightarrow \tilde{\eta}_0$. In each topologically distinct region
the value of $(w_+,w_-)$ is given.
}
\label{evo}
\end{figure*}

\subsection{The phase diagram in the bulk geometry}

Away from the Hermitian limit, the pair of winding numbers $(w_+,w_-)$ specifies topologically distinct regions in $\tilde{\tau}$, each of which represents a gapped topological phase.
In Fig. 1(b), different gapped phases in the bulk geometry
under the mpbc are
shown in the subspace $\{t_1,b\}$ of $\tilde{\tau}$,
and specified by $w_\pm (t_1,b)$.

The phase boundaries of such a winding number map are given by
the gap closings under the mpbc (see Fig.1 as an example).
At the gap closings, the quartic equation (\ref{quartic})
with $E=0$ must hold and gives four solutions of $\beta=be^{ik}$,
which are expressed as $\beta=\beta_\mu$ ($\mu=1,2,3,4$).
This indicates that a solution with $E=0$ appears under the mpbc
if $b=b_\mu (t_1)$ ($\mu=1,2,3,4$), where $b_\mu=|\beta_\mu|$.
That is, each trajectry $b=b_\mu (t_1)$ ($\mu=1,2,3,4$),
on which the gap closes, serves as a phase boundary.
At $E=0$, Eq. (\ref{quartic}) reduces to a set of decoupled 
equations: $R_+ (\beta)=0$ and $R_- (\beta)=0$.
Therefore, two of the four $\beta_\mu$s are
the solutions of $R_+ (\beta)=0$,
and the other two are those of $R_- (\beta)=0$.
Thus, on a phase boundary: $b=b_\mu$, either of the trajectory
$R_\pm (\beta=b_\mu e^{ik})$ passes the origin at $\beta=\beta_\mu$.
Therefore, at this point $w_\pm$ changes by $\pm 1$.

Note that
the gap closings under the mpbc (the phase boundaries: $b=b_\mu$),
generally, do not correspond to any physical reality in the boundary geometry
under the obc
as explained in the next section.

\section{Generalized bulk-boundary correspondence: the prescription}

In the previous section, we have seen
how a pair of winding numbers $(w_+,w_-)$ classifies
different gapped phases in the bulk geometry.
Here, we discuss
how this could be related to the physical reality in the boundary geometry under the obc
(bulk-boundary correspondence).
Below we give a prescription how to establish
a one-to-one relation
between the presence/absence of boundary states under the obc
and
$(w_+,w_-)$ under the mpbc.

\subsection{$P$ and $\tilde{P}$ are corresponding}

To specify a boundary geometry under the obc,
one needs to specify a point $P$ in $\tau$.
At a given point $P$ 
one has a concrete obc spectrum
such as the one represented in Fig. 1 (a) [top annex panel].
At $P$ one can tell from the obc spectrum 
whether the system is in the TI phase: a pair of zero-energy boundary states, 
or in the OI phase: no boundary state.

To define a bulk geometry, one needs to specify also the mpbc parameter $b$; defining a bulk geometry is thus equivalent to specifying a point $\tilde{P}$ in $\tilde{\tau}$ corresponding to $P$.
To establish the bulk-boundary correspondence
the value of $b$ must be specified 
at an arbitrary point $P$ in such a way that 
the boundary property is always in one-to-one correspondence 
with $(w_+, w_-)$ at $\tilde{P}$.

In the obc spectrum shown in Fig. 1 (a) 
one can follow how the boundary property at $P$ evolves as $t_1$ is varied.
In the subspace considered in Fig. 1,
$P=t_1$, 
and the corresponding point $\tilde{P}$ can be denoted as
$\tilde{P}=(t_1,b)$.
The pair of winding numbers $(w_+, w_-)$ at 
an arbitrary point $\tilde{P}$ is given in Fig. 1.
To complete our recipe for achieving the bulk-boundary correspondence,
the function $b=b(t_1)$ must be given
which specifies a path $\tilde{\eta}$
that the point $\tilde{P}$ follows in (the subspace of) $\tilde{\tau}$.

\begin{figure*}
(a)
\includegraphics[width=70mm, bb=0 0 284 284]{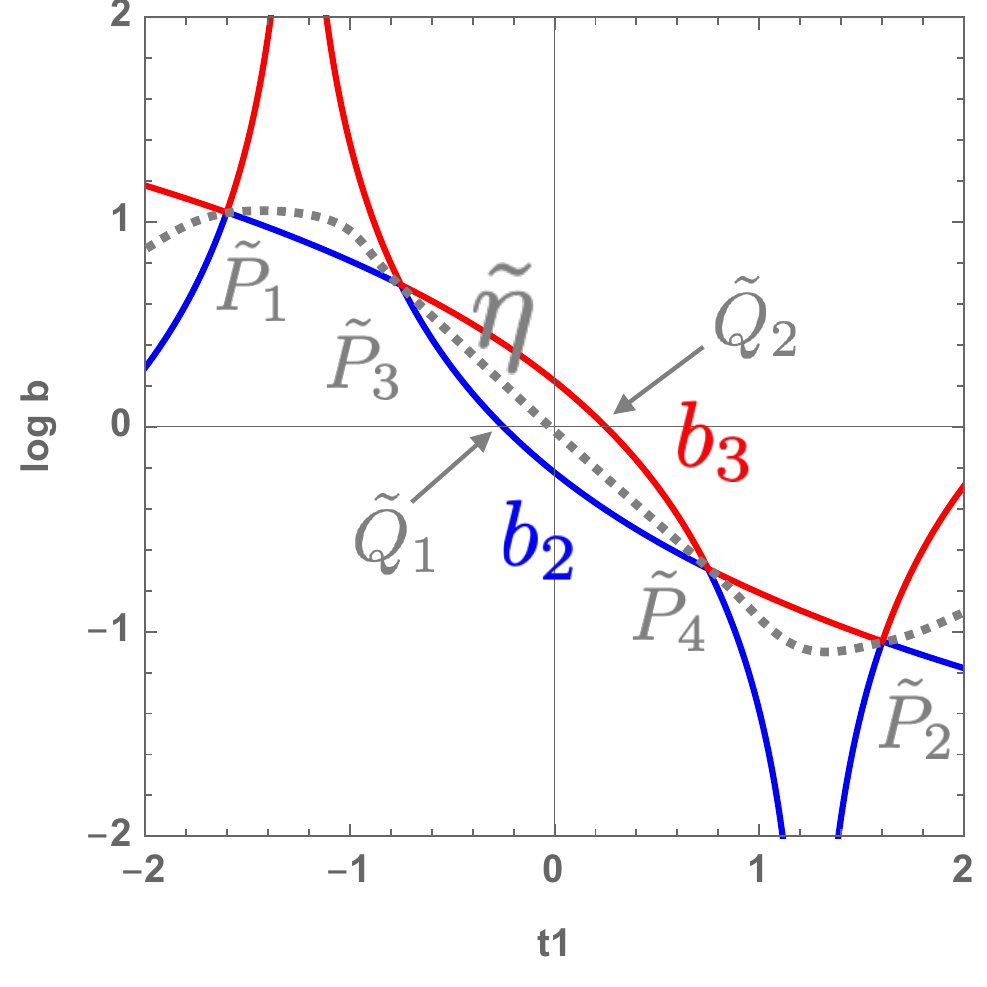}
(b)
\includegraphics[width=70mm, bb=0 0 284 284]{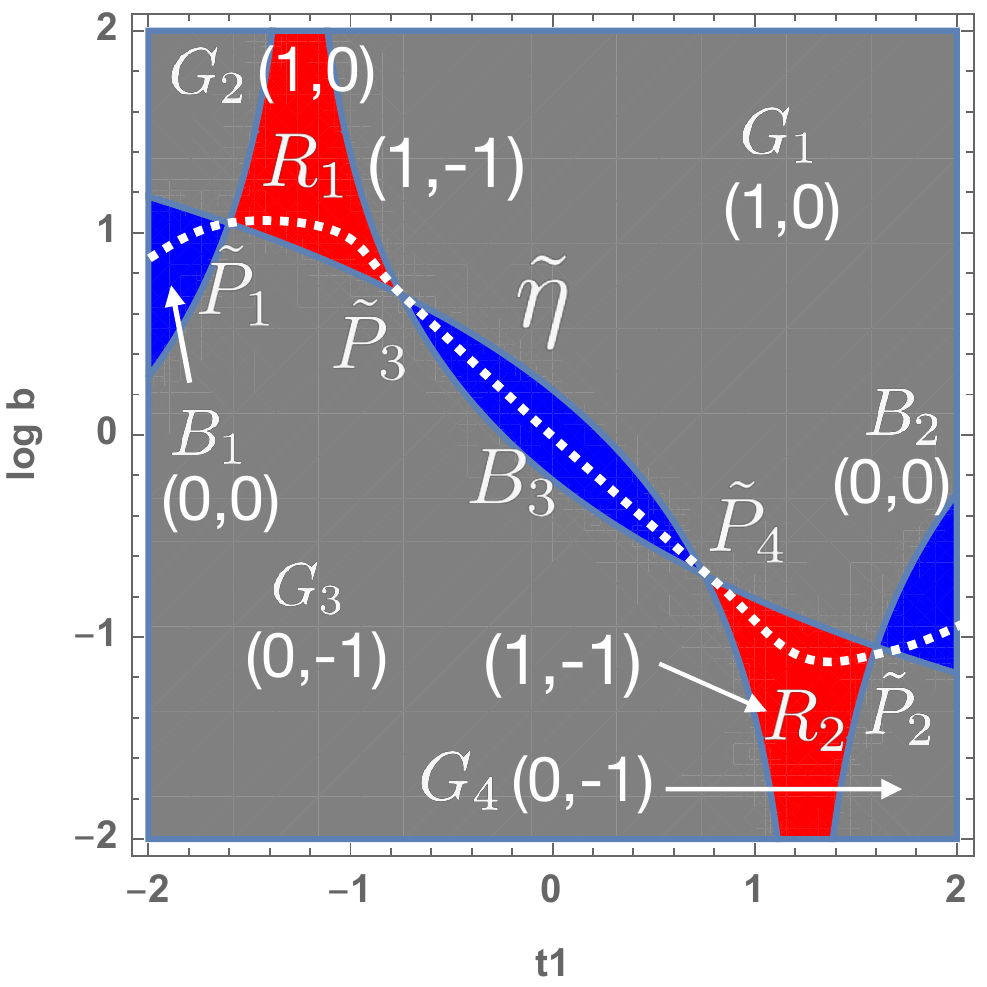}
\vspace{-2mm}
\caption{The trajectories of $b_2$ and $b_3$ [panel (a)]
and the corresponding winding number map [panel (b)]
in the non-perturbative non-Hermitian regime.
Parameters: $t_2=1$, $t_3=0$, $\gamma_1=1.25$, $\gamma_2=0$.
}
\label{nonP}
\end{figure*}

\begin{figure}
\centering\includegraphics[width=80mm, bb=0 0 360 116]{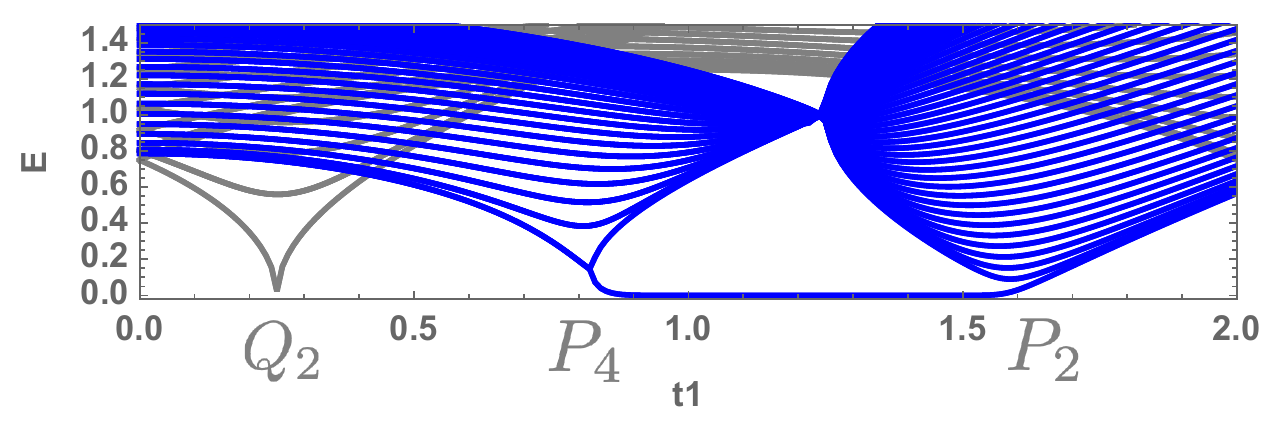}
\vspace{-2mm}
\caption{The obc (pbc) spectrum [shown in blue (gray)] corresponding to Fig. \ref{nonP}.
Only the $t_1>0$ part is shown.
A pair of zero-energy end states appears
between the points $P_4$ and $P_2$, 
corresponding, respectively, to $\tilde{P}_4$ and $\tilde{P}_2$ 
in Fig. \ref{nonP}.
The path $\tilde{\eta}$ is in the TI region $R2$ in this range of $t_1$.
}
\label{spec_nonP}
\end{figure}

\subsection{Perturbative non-Hermitian regime}

In the Hermitian limit: $\gamma_1=\gamma_2=0$,
the bulk-boundary correspondence is done by
smooth deformation of the model parameters
so as not to change the winding number $w$.\cite{ryu}
The model parameters in the TI region with boundary states
can be smoothly reduced to $t_1=0$, $t_2=1$ and $t_3=0$
with keeping $w=1$.
Those in the OI region with no boundary state
can be reduced to $t_1=1$, $t_2=0$ and $t_3=0$
with keeping $w=0$.
In the non-Hermitian case
we do the same kind of smooth deformation in $\tilde{\tau}$.\cite{IT19}

We consider the perturbative non-Hermitian regime,
in which the arrangement of TI and OI phases in the boundary geometry
is identical to the one in the Hermitian limit,
i.e., only the position of the phase boundaries
differs from the Hermitian limit.
Let us define $\tilde{\eta}_0$ as the trajectory of $\tilde{P}$ in the Hermitian limit.
Namely, $\tilde{\eta}_0=(t_1,b=1)$.
An example is shown in Fig. 4(c).
Note that $\tilde{\eta}_0$ goes through all the crossing points of $b_2$ and $b_3$
and apart from these points it stays in a region sandwiched between $b_2$ and $b_3$;
i.e.,
\begin{equation}
b_{2}(t_{1}) < b(t_{1})=1 < b_{3}(t_{1}).
\end{equation}
If $\tilde{\eta}$ is smoothly connected to $\tilde{\eta}_0$,
the bulk-boundary correspondence is at work.
By $\tilde{\eta}$ being smoothly connected to $\tilde{\eta}_0$, we mean that
the arrangement of $(w_+,w_-)$ on $\tilde{\eta}$
must be the same as the ones on $\tilde{\eta}_0$. 

In considering $\tilde{\eta}$, it is useful to note the following.
As $t_1$ varies,
$P=t_1$ experiences different gapped phases (TI, or OI)
in the boundary geometry (see Fig. 1).
Bulk energy gap closes at the phase boundary between neighboring gapped phases.
We name such phase boundaries as $P_j$s ($j=1,2,\cdots$) with $P_{j} = t_{1}^{(j)}$.
Then,
the two trajectories $b_2(t_1)$ and $b_3(t_1)$ cross at $t_{1} = t_{1}^{(j)}$,
and this must be always the case.
The reason is the following.
As the bulk gap closes in the boundary geometry at $P_{j} = t_{1}^{(j)}$,
there must exist a zero-energy solution in the form of Eq. (\ref{psi4}).
This combined with Eq. (\ref{YM_cond}) requires that
Eq. (\ref{quartic}) with $E=0$ gives $\beta_\mu$s satisfying
\begin{equation}
|\beta_{1}(t_{1}^{(j)})| < |\beta_{2}(t_{1}^{(j)})| = |\beta_{3}(t_{1}^{(j)})|
                       < |\beta_{4}(t_{1}^{(j)})|.
\label{sec2}
\end{equation}
Note that the secular equation is independent of a boundary condition.
Combining Eq. (\ref{sec2}) with the definition of $b_{\mu}(t_{1})$
in the bulk geometry (i.e., $b_{\mu}(t_{1}) = |\beta_{\mu}(t_{1})|$
with $\beta_{\mu}(t_{1})$s being solutions of the secular equation at $E = 0$), we readily find $b_{2}(t_{1}^{(j)}) = b_{3}(t_{1}^{(j)})$.
This indicates that $b_2(t_1)$ and $b_3(t_1)$ must cross at $t_{1}=t_{1}^{(j)}$.

As $\tilde{\eta}_0$ goes through all the crossing points of $b_2$ and $b_3$,
$\tilde{\eta}$ must go through all the corresponding points
in order to be smoothly connected to $\tilde{\eta}_0$.
That is, the crossing point of $b_2(t_1)$ and $b_3(t_1)$ at $t_1 = t_1^{(j)}$
should be identified 
as $\tilde{P}_{j}$ corresponding to $P_j$.
Apart from $\tilde{P}_{j}$s,
$\tilde{\eta}$ is not allowed to cross any of the trajectory of $b_\mu$.
If it does, the bulk spectrum becomes gapless,
while the obc spectrum remains gapped
as the bulk condition (\ref{YM_cond}) is unsatisfied
at the corresponding point in $\tau$.

In consequence, the condition for $\tilde{\eta}$ is given as follows:
\begin{enumerate}
\item 
$\tilde{\eta}$ must go through all the crossing points of 
$b_2$ and $b_3$.
\item
Apart from these points
it must stay in a region sandwiched between 
the two trajectories $b_2$ and $b_3$:
\begin{equation}
b_{2}(t_{1}) < b(t_{1}) < b_{3}(t_{1}).
\label{ineq}
\end{equation}
\end{enumerate}
Figure \ref{evo} shows how such a smooth deformation is done in $\tilde{\tau}$.
$\tilde{\eta}_2$ in panel (a)
is smoothly deformed to
$\tilde{\eta}_1$ in panel (b), then to
$\tilde{\eta}_0$
in the Hermitian limit [panel (c)].
Note also that away from the crossing points $\tilde{P}_{j}$,
the choice of $\tilde{\eta}$ is not unique.

In Figs. 1 and 4
the gray regions with $w=1/2$, corresponding to
$(w_+,w_-)=(1,0), (0,-1)$,
are new topologically distinct regions
absent in the Hermitian limit.
The above argument shows that $\tilde{\eta}$ does not pass either of such a region.
Thus, the new topological region with $w=1/2$ will not manifest any physical reality under the obc.

\begin{figure*}
(a)
\includegraphics[width=70mm, bb=0 0 284 284]{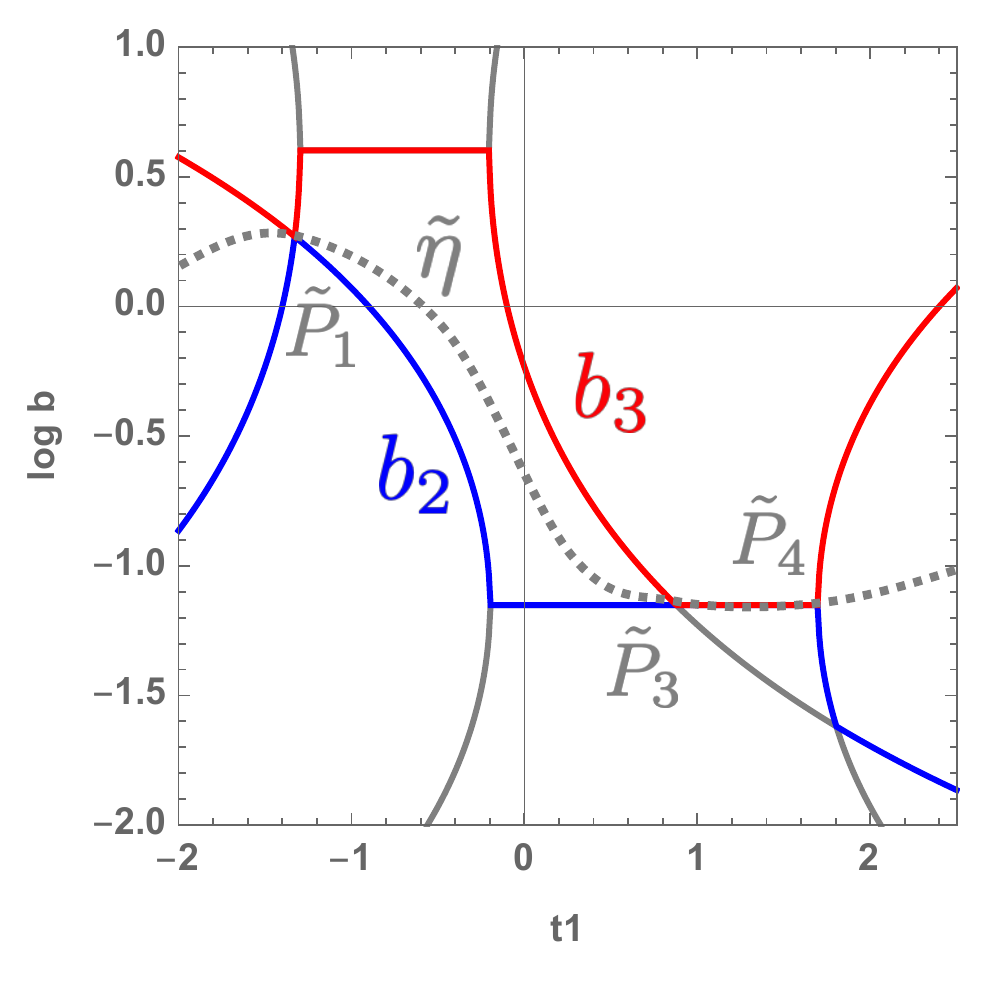}
(b)
\includegraphics[width=70mm, bb=0 0 284 284]{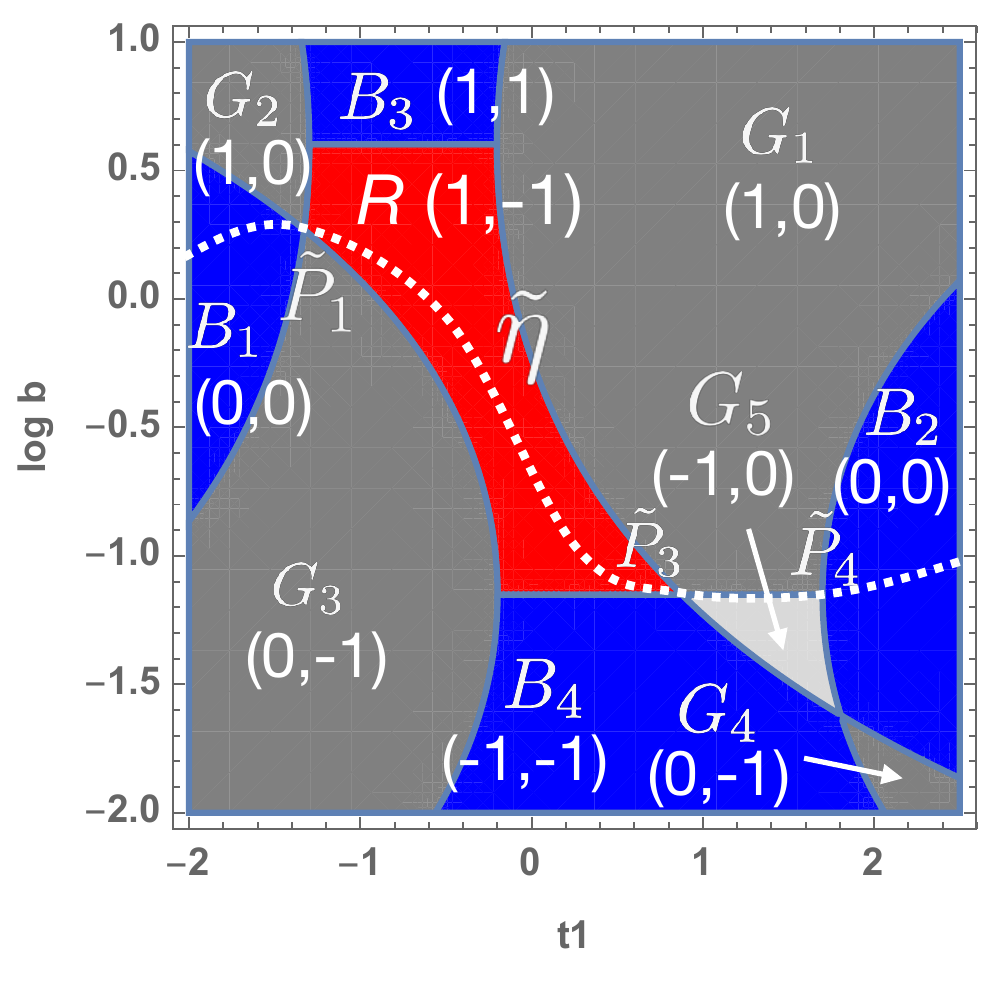}
\vspace{-2mm}
\caption{Case of a finite gapless region (non-perturbative non-Hermitian regime).
Parameters:
$t_2 = 1$, $t_3 = 0.15$, $\gamma_1 = 0.75$,  $\gamma_2 = 0.5$.
}
\label{gapless}
\end{figure*}

\begin{figure}
\centering\includegraphics[width=80mm, bb=0 0 369 128]{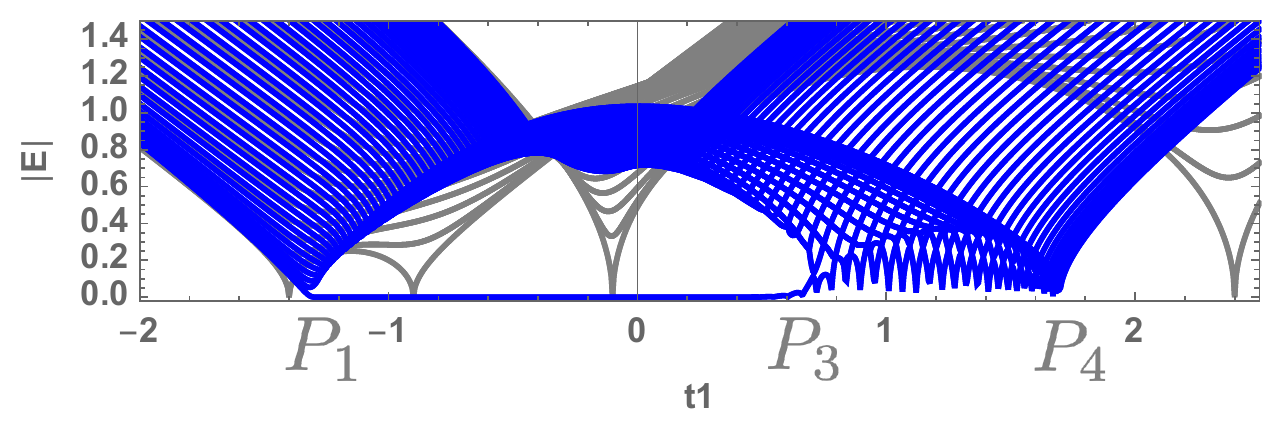}
\vspace{-2mm}
\caption{The obc (pbc) spectrum [shown in blue (gray)] corresponding to Fig. \ref{gapless}.
The bulk part of the
spectrum is expected to be gapless in a finite range of $t_1$
between $P_3$ and $P_4$, corresponding, respectively, to $\tilde{P}_3$ and $\tilde{P}_4$ in Fig. \ref{gapless}.
}
\label{spec_fini}
\end{figure}

In our previous argument
\cite{IT19}
we proposed that
$\tilde{\eta}$ must be chosen to be smoothly connected to $\tilde{\eta}_0$.
Here, we show that
this is equivalent to choosing $\tilde{\eta}$ to satisfy
the inequality (\ref{ineq}).
Below this new recipe is applied to a non-perturbative non-Hermitian regime.

\subsection{Non-perturbative non-Hermitian regime}

Let us next consider the non-perturbative non-Hermitian regime, in which non-Hermiticity is strong enough to induce a new distinct gapped phase absent in the Hermitian limit.
When such a new gapped phase is present in the boundary geometry,
$\tilde{\eta}$ in the bulk geometry cannot be smoothly connected to $\tilde{\eta}_0$.
However,
if one requires the same condition as in the perturbative regime
to $\tilde{\eta}$
[cf. the inequality (\ref{ineq})],
the bulk-boundary correspondence is also at work in the same way.
This is almost trivial if one notices
the continuity of $\tilde{\eta}$ as a curve
before and after the appearance of the new gapped phase.
Note that the change of the arrangement of $(w_+,w_-)$ on $\tilde{\eta}$ becomes discontinuous in this regime.

The phase diagram shown in Fig. \ref{nonP} 
represents an example of such a parameter region.
The appearance of the central OI region $B_3$
with split TI regions $R_1$ and $R_2$
[see panel (b)]
is the new feature absent in the Hermitian limit.
The appearance of $B_3$ leads to
additional crossing points of $b_2$ and $b_3$
($\tilde{P}_3$ and $\tilde{P}_4$).
The inequality (\ref{ineq}) suggests that
$\tilde{\eta}$ goes through these crossing points.
Again, only at these points $\tilde{P}_3$ and $\tilde{P}_4$,
the OI region $B_3$ touches either of the two TI regions.

The path $\tilde{\eta}$ incident in the TI region $R_1$ 
goes through $\tilde{P}_3$,
enters the OI region $B_3$, 
going through $\tilde{P}_4$, 
arrives at the TI region $R_2$,
then follows the original (perturbative) track.
Between $\tilde{P}_3$ and $\tilde{P}_4$, 
$\tilde{\eta}$ must go through 
the OI region $B_3$ without touching or crossing
$b_2$ or $b_3$, but must stay in between.
At any moment, $\tilde{\eta}$ is not allowed to enter
either of the gray region: $G_1$ or $G_3$.

In this case
$\tilde{\eta}$ is not smoothly connected to $\tilde{\eta}_0$,
but the correspondence to the obc spectrum is still maintained.
Figure 6 represents an obc spectrum (shown in blue)
corresponding to Fig. \ref{nonP},
where the $t_1>0$ part of the spectrum is shown.
The gapped region at small $t_1$ corresponds to the OI region $B_3$.
As $t_1$ is increased,
a pair of zero-energy end states appear
between $P_4$ and $P_2$,
corresponding to the TI region $R_2$.
One can thus verify that the additional crossing point $\tilde{P}_4$
corresponds to a new gap closing $P_4$.

\subsection{Finite gapless region}

Finally, let us consider a parameter region employed in Fig. \ref{gapless}.
In this region,
another feature absent in the Hermitian limit
becomes possible:
the degeneracy of $b_2$ and $b_3$ [the condition (\ref{sec2})]
maintained in a finite region of $t_1$
between $\tilde{P}_3$ and $\tilde{P}_4$.
The path $\tilde{\eta}$ incident in the TI region $R$ with $(w_+,w_-)=(1,-1)$
goes through this infinitely narrow path
$\tilde{P}_3\tilde{P}_4$
to enter the OI region $B_2$.
Under the obc (Fig. \ref{spec_fini}) the path $\tilde{P}_3\tilde{P}_4$ corresponds to a 
quasi-gapless region between $P_3$ and $P_4$.\cite{YM}
In the limit $L\rightarrow\infty$,
the spectrum is expected to become gapless in this finite segment $P_3 P_4$.
%
The corresponding path $\tilde{P}_3\tilde{P}_4$ is characterized as
the phase boundary between the half-integral $w$ regions in the bulk geometry:
$G_1$ and $G_5$ (the gray and light-gray regions).
$G_5$ is a newly emerged region with $w=-1/2$, $(w_+,w_-)=(-1,0)$,
distinct from $G_1$ with $w=1/2$, $(w_+,w_-)=(1,0)$.

\section{Conclusions}

Motivated by the newly proposed non-Hermitian topological insulator,
we have developed non-Hermitian Bloch band theory. 
The concept of 
bulk-boundary correspondence is 
the central idea of topological insulator,
while in some non-Hermitian models with non-reciprocal hopping
this correspondence is superficially broken.
In such non-Hermitian systems
due to the non-Hermitian skin effect,
the bulk spectrum is somewhat too sensitive to a boundary condition
for the standard procedure to be valid as such.

We have proposed to employ the modified periodic boundary condition
(mpbc) for describing the bulk of such a system.
The use of this mpbc
appropriately captures 
the features specific to non-Hermitian systems,
and underlies
the Bloch band theory for non-Hermitian systems.
We have shown that 
the bulk-boundary correspondence is
restored in the generalized parameter space $\tilde{\tau}=\{\tau,b\}$,
where
$\tau$ represents the original model parameter space,
and $b$ specifies the mpbc.

In the generalized bulk-boundary correspondence,
a pair of chiral winding numbers $(w_+,w_-)$ 
evaluated on the path $\tilde{\eta}$ in $\tilde{\tau}$
plays the role of characterizing each topologically distinct region.
We present a simple recipe to give $\tilde{\eta}$ in a proper manner.

To illustrate our scenario 
we have employed a non-Hermitian version of the SSH model.
However, the proposed scenario itself should be applicable to a broader class
of non-Hermitian topological insulators.
A relation analogous to Eq. (\ref{ineq}) can be equally found for 
a different class of models, e.g.,
for models with symplectic symmetry. \cite{KK_PRB2020,Okuma_PRL}
In such cases the concrete relation
(\ref{ineq})
should be replaced with a suitable one, but the rest of the recipe is unchanged.
Restoring the bulk-boundary correspondence in its proper sense 
as a correspondence between the genuine bulk and boundary quantities, 
our generalized Bloch band theory
set the basis on which
the concept of topological insulator is firmly 
established in non-Hermitian systems.

\acknowledgments
The authors thank K. Shobe, K. Kawabata, N. Hatano, H. Obuse,
Y. Asano, Z. Wang, C. Fang and T. Ohtsuki
for discussions and correspondences. 
This work has been supported by JSPS KAKENHI 
Grants 
No. 15H03700,
No. 15K05131,  
No. 18H03683,
No. 18K03460,
and
No. 20K03788.

\bibliography{ptep4}

\end{document}